\documentclass[a4paper,fleqn,usenatbib]{mnras}

\usepackage{newtxtext,newtxmath}

\usepackage[T1]{fontenc}
\usepackage{ae,aecompl}

\usepackage{graphicx}	
\usepackage{amsmath}	
\usepackage{amssymb}	
\usepackage{subfigure}
\usepackage{overpic}
\usepackage{pdflscape}
\usepackage{multicol}        
\usepackage{bm}		

\title[GLF of BL Lacs and contribution to EGB ]{Gamma-ray luminosity function of BL Lac objects and contribution to the extragalactic gamma-ray background }

\author[Qu et al.]{
Yankun Qu$^{1,2}$
Houdun Zeng,$^{1}$ \thanks{E-mail: zhd@pmo.ac.cn}
Dahai Yan$^{3}$ \thanks{E-mail: yandahai@ynao.ac.cn}
\\
$^{1}$Key laboratory of Dark Matter and Space Astronomy, Purple Mountain Observatory, Chinese Academy of Sciences , Nanjing 210008,China\\
$^{2}$University of Chinese Academy of Sciences, Yuquan Road 19, Beijing, 100049, China\\
$^{3}$Key Laboratory for the Structure and Evolution of Celestial Objects, Yunnan Observatory, Chinese Academy of Sciences, Kunming 650011, China
}

\date{Accepted XXX. Received YYY; in original form ZZZ}

\pubyear{2015}

\begin{document}
\label{firstpage}
\pagerange{\pageref{firstpage}--\pageref{lastpage}}
\maketitle

\begin{abstract}
Using a significantly enlarged \textit{Fermi}-LAT BL Lac objects (BL Lacs) sample, we construct the gamma-ray luminosity function (GLF) of BL Lacs,
 by the joint use of the space density distribution and source counts distribution.
 We use three well-studied forms of the GLF, i.e., the forms of pure density evolution (PDE), pure luminosity
evolution (PLE) and luminosity-dependent density evolution (LDDE).
The Markov Chain Monte Carlo (MCMC) technique is used to constrain model parameters.
Our results suggest that LDDE model can give the best description for the BL Lac GLF. And the model shows that the BL Lacs with a harder GeV spectrum and a less luminosity evolve as strongly as FSRQs,
and the evolution decreases as increasing luminosity.
We also model the average photon spectra of BL Lacs with a double power laws model.
Using this modeled spectra, BL Lacs contribute $ \sim 20\%$ of the total extragalactic gamma-ray background (EGB) at $E>100$ MeV,
$\sim 100\%$ of the EGB at $E>50$ GeV, and the unresolved BL Lacs contribute $\sim 20 \%$ of the isotropic diffuse gamma-ray background (IGRB) at $E>100$ MeV.
A prediction of the TeV EGB spectra are given, which may be tested by the future detectors.
\end{abstract}




\section{Introduction}

One of the fundamental astrophysics problems is the origin of the extragalactic gamma-ray background (EGB) \citep{1995ApJ...452..156C,2010PhRvL.104j1101A,Abdo_2010},
which was first detected by the SAS-2 mission \citep{1975ApJ...198..163F}.
Later the Energetic Gamma-Ray Experiment Telescope (EGRET) on board the Compton Observatory
proved  that the  spectrum of EGB was a featureless power law with a photon index of 2.4 in the 0.2$-$100 GeV band \citep{2004ApJ...613..956S,1998ApJ...494..523S,1996ApJ...464..600S}.
In the EGRET era, the observed EGB intensity above 100 MeV  is $ 1.45(\pm0.05)\times10^{-5} \rm ph \, cm^{-2}  s^{-1}  sr^{-1}$ \citep{1998ApJ...494..523S}.
\textit{Fermi}-LAT has made a new
measurement of the EGB spectrum, and the results show that the EGB energy spectrum between 0.1 and 820 GeV
can be well represented by a power law with an exponential cutoff above $\sim $300 GeV \citep{Ackermann_2015}.
The total EGB and  the isotropic diffuse gamma-ray background
(IGRB) intensity above 100 MeV are $1.13(\pm 0.17) \times 10^{-5} $ ph  cm$^{-2}$  s$^{-1}$ sr$^{-1} $ and $7.2(\pm 0.6) \times 10^{-6}$ ph  cm$^{-2}$  s$^{-1}$  sr$^{-1} $ respectively, significantly lower than the EGRET's result. \cite{PhysRevLett.116.151105} suggested that the hard \textit{Fermi}-LAT sources (2FHL) can account for the EGB above 50 GeV, and this catalog 2FHL is mainly composed of the BL Lacertae objects (BL Lacs) population.

The EGB is considered to be the superposition of contributions from diffuse emission processes such as the decay
 or annihilation of dark matter \cite[e.g.][]{1996PhR...267..195J,2002PhRvD..66l3502U} and the interactions of cosmic rays with background photons \cite[e.g.][]{2009PhRvD..79f3005K,2011PhRvD..84h5019A},
 and unresolved extragalactic sources including active galactic nuclei (AGNs)\cite[e.g.][]{2012ApJ...751..108A, 2015ApJ...800L..27A, 2014JCAP...11..021D,2014ApJ...780..161D,2014ApJ...786..129D,2015PhRvD..91l3001D}, starburst galaxies \cite[e.g.][]{2012ApJ...755..164A,2014ApJ...793..131C} and gamma-ray bursts (GRBs) \cite[e.g.][]{2007ApJ...656..306C}.
 Among these contributors, AGNs play an important role, especially at $E>100$ GeV energies.
 Most of \textit{Fermi}-LAT AGNs are blazars, which include two main subgroups: BL Lacs and flat  spectrum radio quasars (FSRQs).
 Based on the second \textit{Fermi}-LAT AGN catalog (2LAC), the gamma-ray luminosity functions (GLFs) of BL Lacs and FSRQs have been constructed
 \citep[e.g.][]{2012ApJ...751..108A, Ajello_2013, 2014ApJ...786..129D,2013MNRAS.431..997Z,zeng2014}.
 Similar to the population of X-ray-selected radio-quiet AGNs,
 gamma-ray FSRQs show a positive cosmological evolution \citep[e.g.][]{2012ApJ...751..108A}.
 BL Lacs present more complicated cosmological evolution \citep{Ajello_2013}.
The evolution of most BL Lacs is positive with a space density peaking at modest
redshift and low luminosity, while the high-synchrotron-peaked (HSP) BL Lacs show strong negative evolution
with number density increasing for $z \leq 0.5$ \citep{Ajello_2013}. \cite{2014ApJ...786..129D} built the GLF models using the 148
BL Lacs in the 2FGL, which have well measured redshifts and SED classification. Their results indicated that the luminosity-dependent density evolution
(LDDE) model is preferred over the pure luminosity evolution
(PLE) and the steep-spectrum radio source (SSRS) models in
reproducing the whole BL Lac population.
 \citet{zeng2014} also studied the the GLF of BL Lacs by using the sample of 175 BL Lacs with known redshifts in \textit{Fermi}-LAT 2LAC,
 and found that LDDE model is better than the pure PLE model and pure density evolution (PDE) model,
 which is in line with \cite{Ajello_2013} and \cite{2014ApJ...786..129D}.

 A further study on cosmological evolution of BL Lacs is indispensable to better understand the origin of the EGB.
In the third \textit{Fermi}-LAT AGN catalog (3LAC), 604 BL Lacs are identified, which is approximately considered to be a flux-limited \textit{Fermi} sample with $S_{25} \succeq 3.0 \times 10^{-12}$ erg cm$^{-2}$ s$^{-1}$ \citep{2015ApJ...810...14A}, among which 307 BL Lacs' redshift are measured. The redshift completeness for BL Lacs is about $50.8\%$. In this work, the fourth \textit{Fermi}-LAT AGN catalog (4LAC) is used to revisit the GLF of BL Lacs and the contribution from BL Lacs to the EGB.
 The \textit{Fermi} 4LAC, which is the results obtained from the first 8-years \textit{Fermi}-LAT data, includes 2863 AGNs, increasing by 80\% over the 3LAC \citep{2019arXiv190510771T}.
 650 FSRQs, 1052 BL Lacs, 1092 BCUs, and 68 other AGNs are included in \textit{Fermi} 4LAC, and about 1347 sources are not reported in \textit{Fermi} 3LAC.
 The enlarged sample of BL Lacs could improve the construction of GLF.

This paper is organized as follows.
 In Section 2, we describe our used sample.
 The method and results for constructing the GLF of BL Lacs are shown in Section 3.
 In Section 4, the contribution of BL Lacs to the EGB and IGRB is estimated based on the our GLF.
 Finally, a brief conclusion and discussion is given in Section 5.
 Throughout this paper we adopt a cold dark matter universe with the matter density parameter $\Omega_m$ = 0.315,  $\Omega_{\Lambda}$= 0.685, and Hubble constant $H_{0} = 67.36$ km s$^{-1}$ Mpc$^{-1}$ \citep{2014A&A...571A..31P}.

\section{Samples}
Very recently, the Fourth {\it Fermi} Large Area Telescope Catalog (4FGL;
\cite{2019arXiv190210045T}) report on 5065 sources with a Test Statistic (TS) value greater
than 25.
Most of them are
active galaxies of the blazar class. Based on 4FGL,
\cite{2019arXiv190510771T} present the
fourth catalog of active galactic nuclei (AGNs) detected by the {\it Fermi}-LAT (4LAC),
containing 2863 AGNs of various
types located at high Galactic latitude, i.e. $|b|>10^{\circ}$.
After removing the entries in 4LAC for which the corresponding
gamma-ray sources were not associated with AGNs and that has more than one counterpart or are
tagged for other reasons in the analysis, \cite{2019arXiv190510771T} obtains a ``clean'' sample of 2649 sources, including
598 FSRQs, 1018 BL Lacs, 972 BCUs and 61 non-blazar AGNs.
The energy flux distribution of all the {\it Fermi} sources
can be seen in Figure~9 of \cite{2019arXiv190210045T}.
The flux threshold in 4FGL is
$\simeq 2 \times 10^{-12}$ erg cm$^{-2}$ s$^{-1}$, lower than the value $\simeq 3 \times 10^{-12}$ erg cm$^{-2}$ s$^{-1}$ in 3FGL.
Note that the redshift distributions of 3LAC and newly detected
blazars have similar means and widths.

Among the 1018 BL Lacs in 4LAC,  649 BL Lacs' redshift are well measured\footnote{Note that 4FGL J0601.3-7238 (PWN J0601-7238) and 4FGL J0719-4012 (1RXS J071939-401153) are not included due to no redshift information in NASA/IPAC Extragalactic Database (NED) and the redshift of 4FGL J0828.3-4152 (B3 0824+420) is 0.223,
which can be found in NED \citep{1999A&AS..139..575W}}.
The redshift completeness of the BL Lacs in 4LAC is about $ 64 \%$, which is improved relative to 3LAC ($51\%$).
The sample of the 649 BL Lacs is used to construct new GLF.

\section{Gamma-ray Luminosity Function}
We build the GLFs of BL Lacs by using the above sample.
Here the GeV gamma-ray spectrum of each BL Lac is assumed to be a simple power-law form with photon index $\Gamma$,
i.e, $F_{\gamma}(E)= N_0 E^{-\Gamma}$. We can obtain the gamma-ray luminosity $L_{\gamma}$  by \citep[e.g.][]{2009MNRAS.396L.105G}
\begin{equation}
\label{Eq.2}
L_{\gamma}=4 \pi d_{L}^{2} S_{\rm obs}  K ,
\end{equation}
where $d_{L}$ is the luminosity distance, $S_{\rm obs}$ is the integrated flux between 100 MeV and 100 GeV, and $K=(1+z)^{\Gamma-2}$ is the $K$-correction term for the observed fluxes into the rest frames.


GLF is defined as the number of sources with $\gamma$-ray luminosities in the range of $L$ to $ L + dL$ per unit of the comoving volume element.
The space density of BL Lacs as the function of $L_{\gamma}$, $z$, and $\Gamma$
is expressed as \citep[e.g.][]{2012ApJ...751..108A, Ajello_2013}
\begin{equation}
\label{eq:SD}
\frac{d^3 N}{dz\,dL_{\gamma}\,d\Gamma} =  \frac{d^2 N}{dL_\gamma\,dV_{\rm com}} \times \frac{dN}{d\Gamma} \times
\frac{dV_{com}}{dz} = \Psi(z,L_\gamma) \times \frac{dN}{d\Gamma} \times
\frac{dV_{\rm com}}{dz}\;,
\end{equation}
where $\Psi(z,L_\gamma)$ is the GLF, $dN/d\Gamma$ is the intrinsic distribution of photon indices,
and $dV_{\rm com}/dz$ is the comoving volume element per unit redshift and unit solid angle.
We adopt the mathematical forms of the GLF given by \cite{Ajello_2013}, 
which are primarily luminosity evolution (PLE), primarily density evolution (PDE), and luminosity-dependent density evolution (LDDE).

Using the space density, we write the probability distribution as
\begin{equation}
\label{eq:Pd}
p(L_{\gamma}, z, \Gamma)=\frac{1}{N_{\rm exp}}\frac{d^3 N}{dz\,dL_{\gamma}\,d\Gamma} \omega(F_{\gamma}),
\end{equation}
where $N_{\rm exp}$ is the expected number of source detections, and $\omega(F_{\gamma})=\omega(L_{\gamma}(F_{\gamma}), z, \Gamma)$ is
the detection efficiency which represents the probability of detecting a BL Lacs with $L_{\gamma}$ and $\Gamma$ at redshift $z$. 

The likelihood function for the observed data $p(L_{\gamma}, z, \Gamma\mid\theta)$
can be derived, once we assume a parametric form for the space density
with parameters $\theta$. \citet{1983ApJ...269...35M} gave a likelihood
function based on the Poisson distribution and defined
$S = - 2 \textrm{ln} (\prod p (L_{\gamma}, z, \Gamma\mid\theta))$ \citep[e.g.][]{1995ApJ...452..156C,2007Ap&SS.309...73N,2009ApJ...699..603A,2016ApJ...829...95Y}.
Dropping the terms independent of
the model parameters, one finds
\begin{equation}
\label{eq:chi_S}
S=-2\sum_{i}^{N_{\rm obs}} \textrm{ln} \frac{d^3 N}{dz\,dL_{\gamma}\,d\Gamma}+2 N_{\rm obs} \textrm{ln}(N_{\rm exp}),
\end{equation}
where $N_{\rm exp}$ is expressed as
\begin{eqnarray}
\label{eq:Nexp}
N_{\rm exp}
= \Omega \int^{\Gamma_{\rm max}}_{\Gamma_{\rm min}}
\int^{z_{\rm max}}_{z_{\rm min}} \int^{L_{\gamma,{\rm max}}}_{L_{\rm min}}
\frac{d^3 N}{dz\,dL_{\gamma}\,d\Gamma } \omega(F_{\gamma}) dz\,d\Gamma\, dL_{\gamma},
\end{eqnarray}
where $\Omega=10.28 \; \textrm{sr}$ is the
the total sky coverage, which is $|b| \geqslant 10^{\circ}$ in our work. We adopt $\Gamma_{\rm min}=1.4$, $\Gamma_{\rm max}=3.0$, $z_{\rm min}=0.0$, $z_{\rm max}=6.0$, $L_{\gamma,{\rm min}}=4 \times 10^{40}$ erg/s and $L_{\gamma,{\rm max}}=10^{50}$ erg/s.

\subsection{Detection efficiency}
\label{sde}
\begin{figure}
\centering
\includegraphics[scale=0.5]{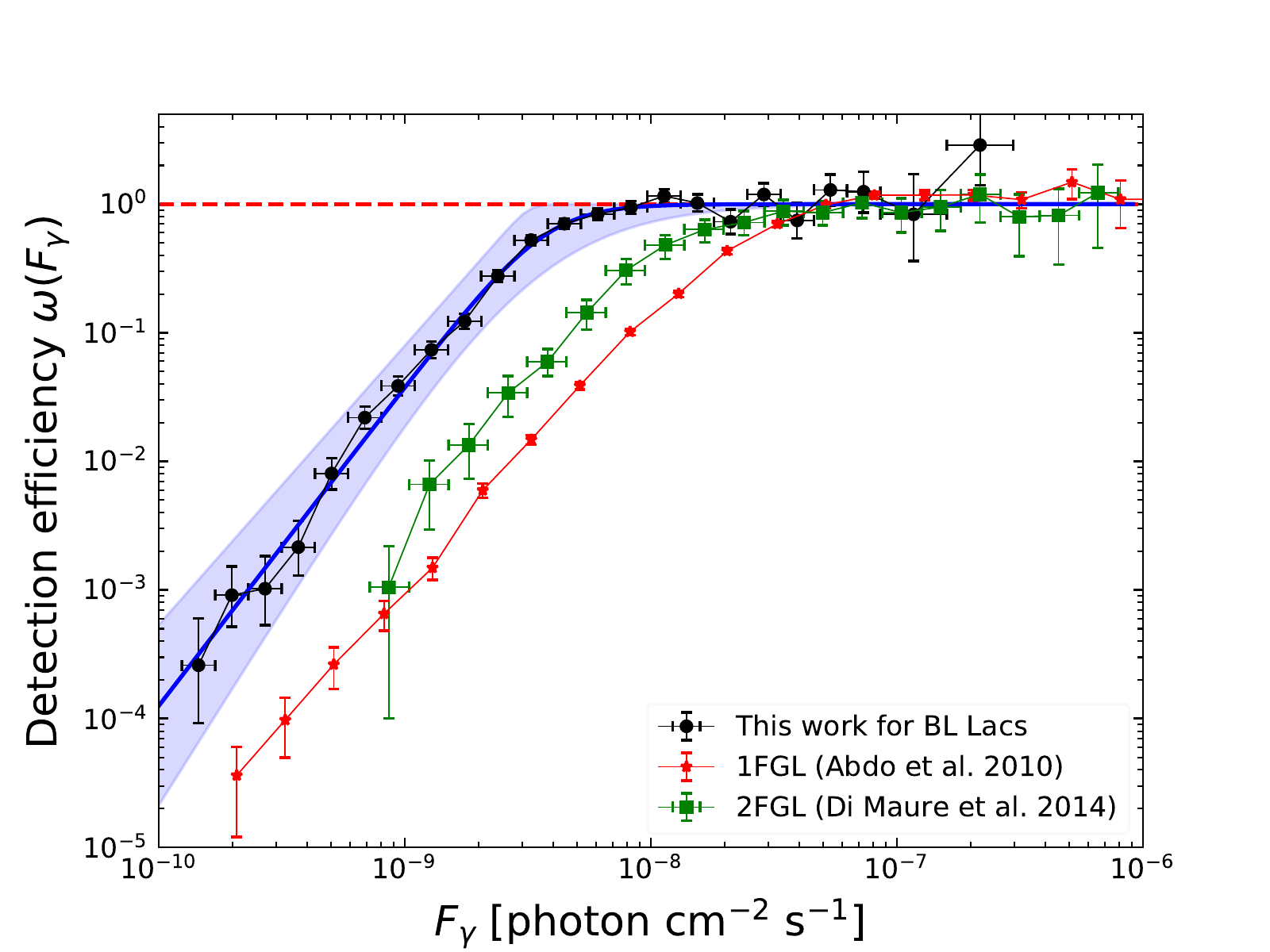}
\centering
\caption{The evaluated detection efficiency of BL Lacs we used in this paper and the evaluated detection efficiency for the 1FGL \citep{Abdo_2010} and the 2FGL \citep{2014ApJ...780..161D} samples.}
\label{de}
\end{figure}

Detection efficiency is an important factor in constructing LF.
According to \citet{Abdo_2010}, \citep{2014ApJ...780..161D} and \citet{Di_Mauro_2018}, the detection efficiency of \textit{Fermi}-LAT can be expressed as
\begin{equation}
\omega(F_{\gamma})=\frac{(dN/dF_{\gamma})_{\rm obs}}{(dN/dF_{\gamma})_{\rm theory}}
\end{equation}
where $(dN/dF_{\gamma})_{\rm obs} = \frac{N_i}{\Omega \vartriangle F_{\gamma,i}}$, $\vartriangle F_{\gamma,i}$ is the width
of flux bin $i$, $N_i$ is the number of sources in our sample, and $(dN/dF_{\gamma})_{\rm theory}$ is the theoretical value.
\citet{Abdo_2010} reported that the $(dN/dF_{\gamma})_{\rm theory}$ is a broken power-law.
And we can obtain result a consistent with their's, the indexes $\beta_1=2.79 \pm 0.46$, $\beta_2=2.08 \pm 0.19$ and the flux break $F_{b}=6.98 \pm 0.37 \times 10^{-8}$ ph cm$^{-2}$ s$^{-1}$, by fitting the observed data with $F_{\gamma}>1.0 \times 10^{-8}$ ph cm$^{-2}$ s$^{-1}$. Figure \ref{de} shows the detection efficiencies we evaluated for our sample, the 1FGL sample in \cite{Abdo_2010} and the 2FGL sample in \cite{2014ApJ...780..161D}. This method is similar to that of \cite{2014ApJ...780..161D}, the details are discussed in the appendix of \cite{2014ApJ...780..161D}. The differences between the
three detection efficiencies at low fluxes are due to the different
observed $dN/dF_{\gamma}$ for the three samples. Since the evaluated  $(dN/dF_{\gamma})_{\rm theory}$ could be different from the real intrinsic distribution, we estimate the 1 $\sigma$ error of $\omega(F_{\gamma})$ at low-end fluxes for our sample, which will bring
systematic errors to the estimation of model parameters.

\subsection{Source counts distribution}
\label{sec:maths} 
In addition, we use the source counts distribution to provide additional constraints on the GLFs.
Note that this source counts distribution is for the whole sample of 1018 BL Lacs. The one-dimensional source counts distribution function is evaluated as
\begin{equation}
\chi ^2=\sum_{i}^{N}(\frac{f_{\rm data,i}-f_{\rm mod,i}}{\sigma_{\rm data,i}})^2\;,
\label{chi}
\end{equation}
where $f = N(> F_{\gamma})$ is the source counts distribution,
and the theoretical source count distribution $f_{\rm mod,i}$ can be written as
\begin{equation}
N(>F_{\gamma}) = \Omega \int^{\Gamma_{\rm max}}_{\Gamma_{\rm min}}
\int^{z_{\rm max}}_{z_{\rm min}} \int^{L_{\gamma,max}}_{L_{\rm limit}(\acute{z},\Gamma,F_{\gamma})}
\frac{d^3 N}{dz\,dL_{\gamma}\,d\Gamma} \omega(F_{\gamma}) dz\,d\Gamma\, dL_{\gamma}\;.
\end{equation}

Ultimately, we can define a new function $S_{\rm all}$ that combines the constraints from the source count distribution and the likelihood function
\begin{equation}
S_{\rm all}=\chi^2+S.
\end{equation}
The fit to the source count distribution enables us to determine the normalization constant of GLF, $A$,  and to better constrain the GLF free parameters. This additional constraint also can be found for RLF of AGN in \cite{2001MNRAS.322..536W} and \cite{2017ApJ...846...78Y}, and for LF of GRB in \cite{2019MNRAS.tmp.1951L}.

The MCMC method \citep[e.g.][]{2010A&A...517L...4F,2011ApJ...735..120Y,2012PhRvD..85d3507L,2013ApJ...765..122Y,Arjas2015Markov} is used to minimize  $S_{\rm all}$
and constrain the model parameters.
The PDE and PLE models have a total of 10
free parameters, i.e., ($A$, $\gamma_1$ , $L_{\ast}$ , $\gamma_2$ , $k$, $\tau$ , $\xi$,  $\mu^{\ast}$, $\beta$, and $\sigma$ ),
and the LDDE model has a total 12 free parameters, i.e., ($A$, $\gamma_1$ , $L_{\ast}$ , $\gamma_2$ , $z^{\ast}_c$, $p1^{\ast}$, $\tau$, 
$p2$, $\alpha$,  $\mu^{\ast}$, $\beta$, and $\sigma$).
Note that the parameter $\mu$ is changed with luminosity as $\mu(L_{\gamma})=\mu^*+\beta\times[ {\rm log}_{10}(L_{\gamma})-46]$,
where $\sigma$ and $\mu$ are the dispersion and the mean of the Gaussian distribution, respectively.
A detailed description of those models can be found in \citet{Ajello_2013}.

\begin{figure*}
\centering
\subfigure[PDE]{
\begin{minipage}[t]{0.48\linewidth}
\includegraphics[width=1.0\linewidth]{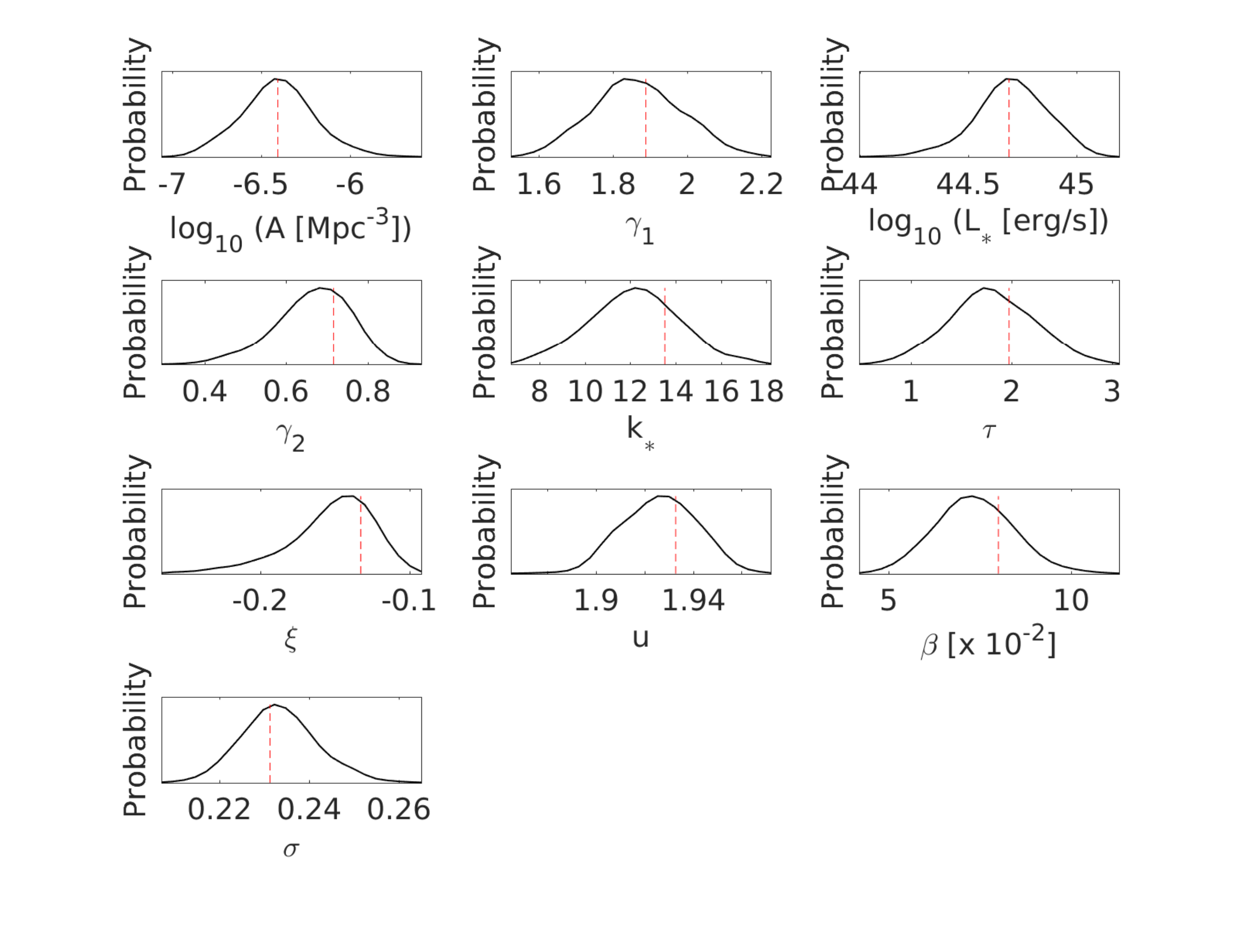}
\end{minipage}
}
\subfigure[PDE]{
\begin{minipage}[t]{0.48\linewidth}
\includegraphics[width=1.0\linewidth]{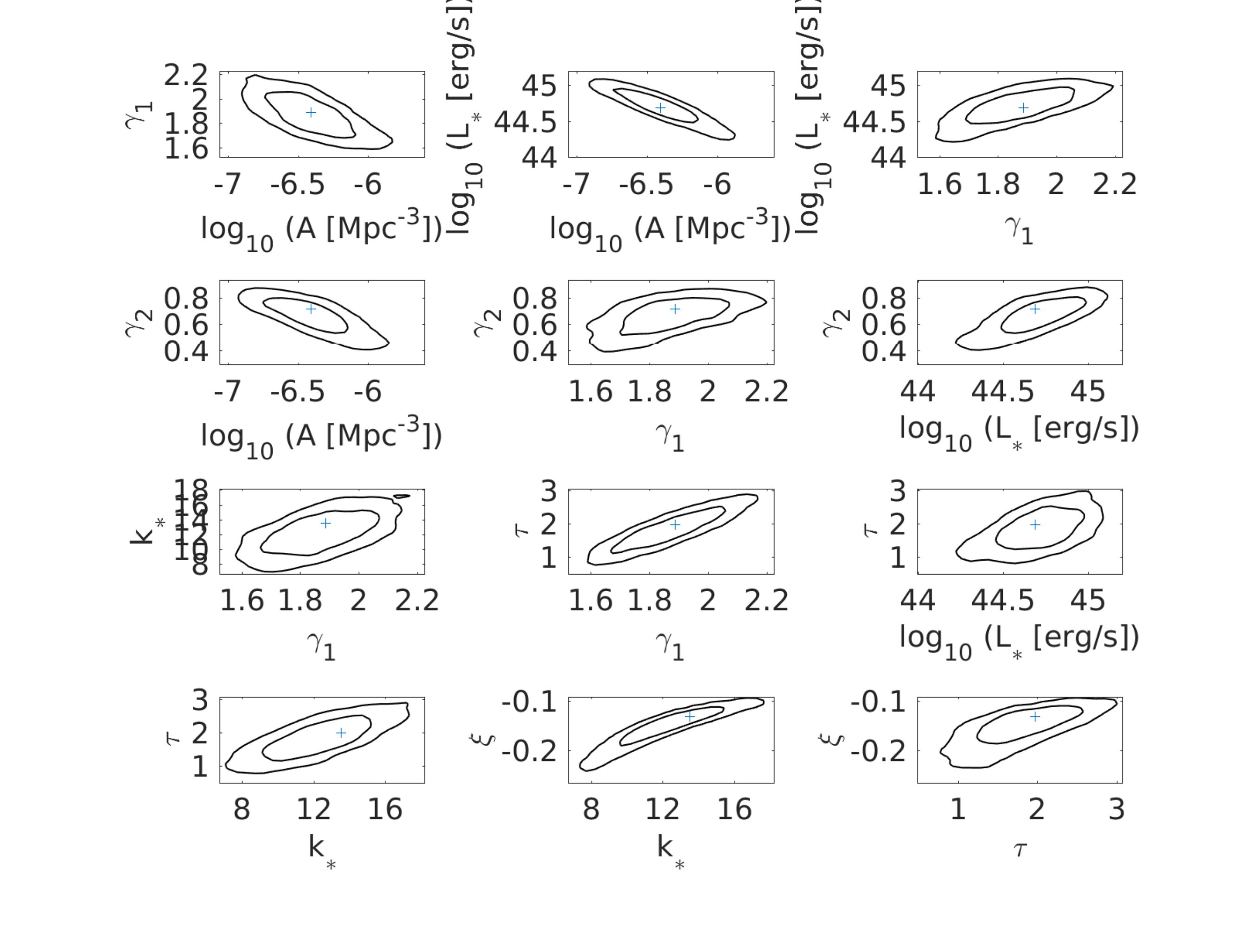}
\end{minipage}
}
\subfigure[PLE]{
\begin{minipage}[t]{0.48\linewidth}
\includegraphics[width=1.0\linewidth]{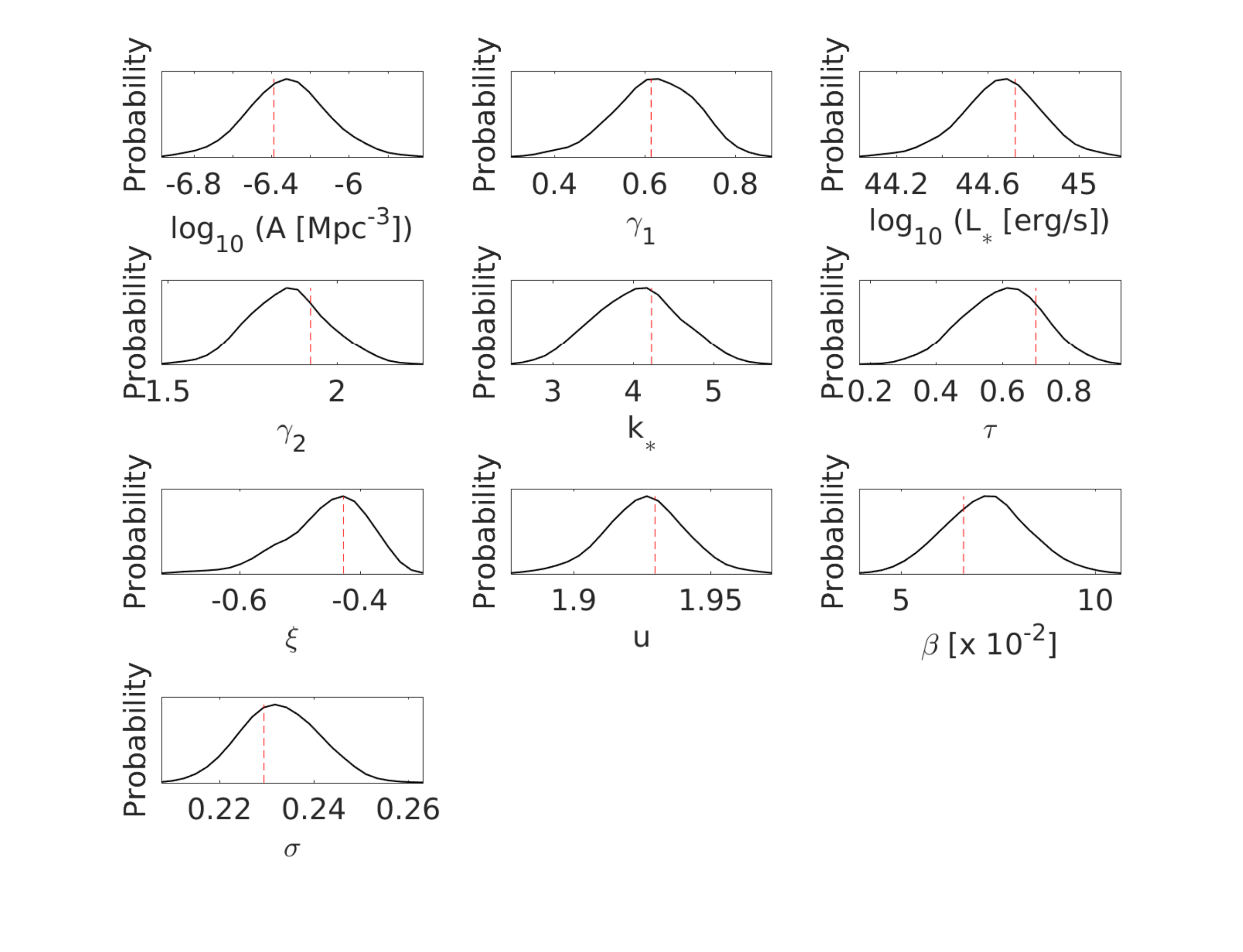}
\end{minipage}
}
\subfigure[PLE]{
\begin{minipage}[t]{0.48\linewidth}
\includegraphics[width=1.0\linewidth]{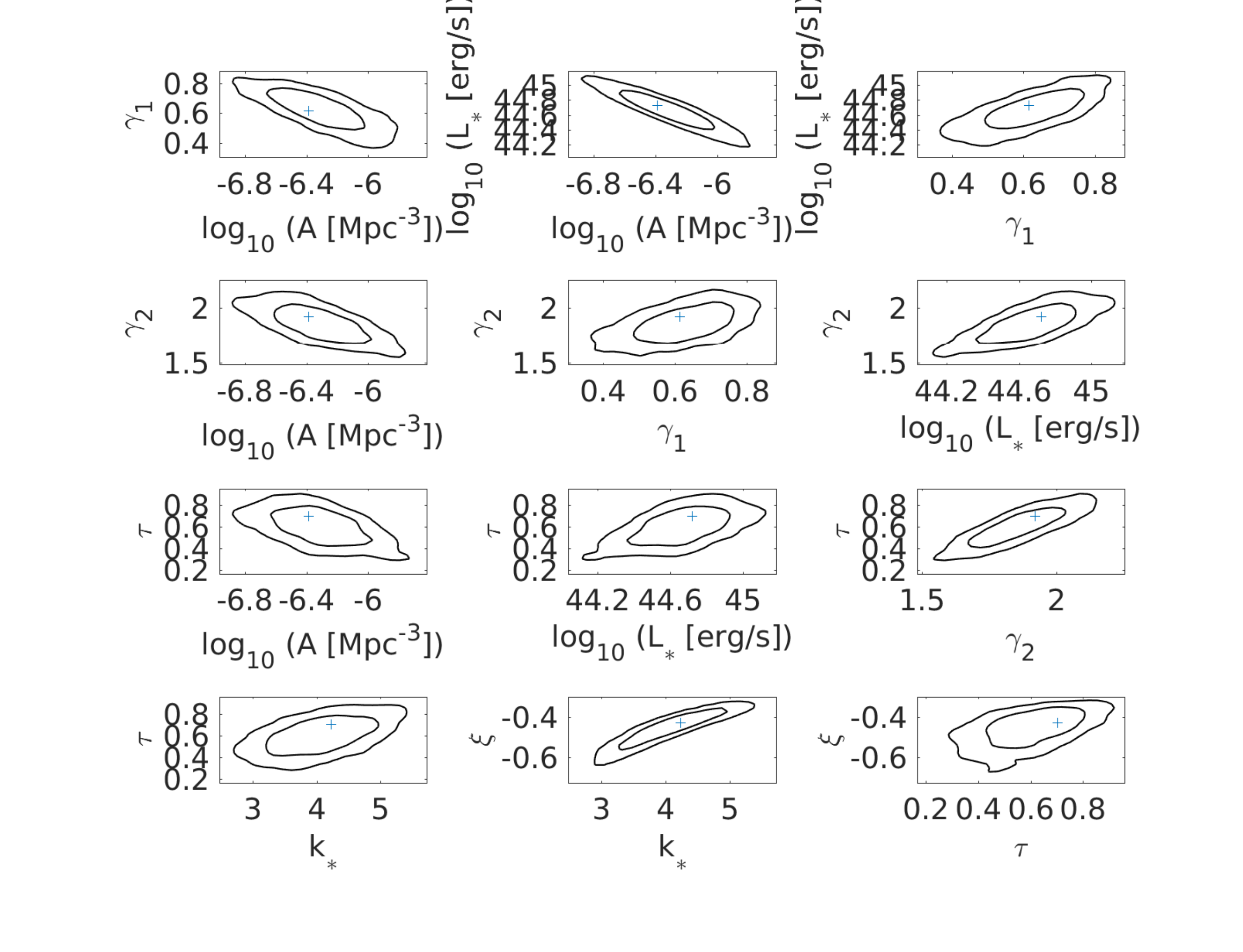}
\end{minipage}
}
\subfigure[LDDE]{
\begin{minipage}[t]{0.48\linewidth}
\centering
\includegraphics[width=1.0\linewidth]{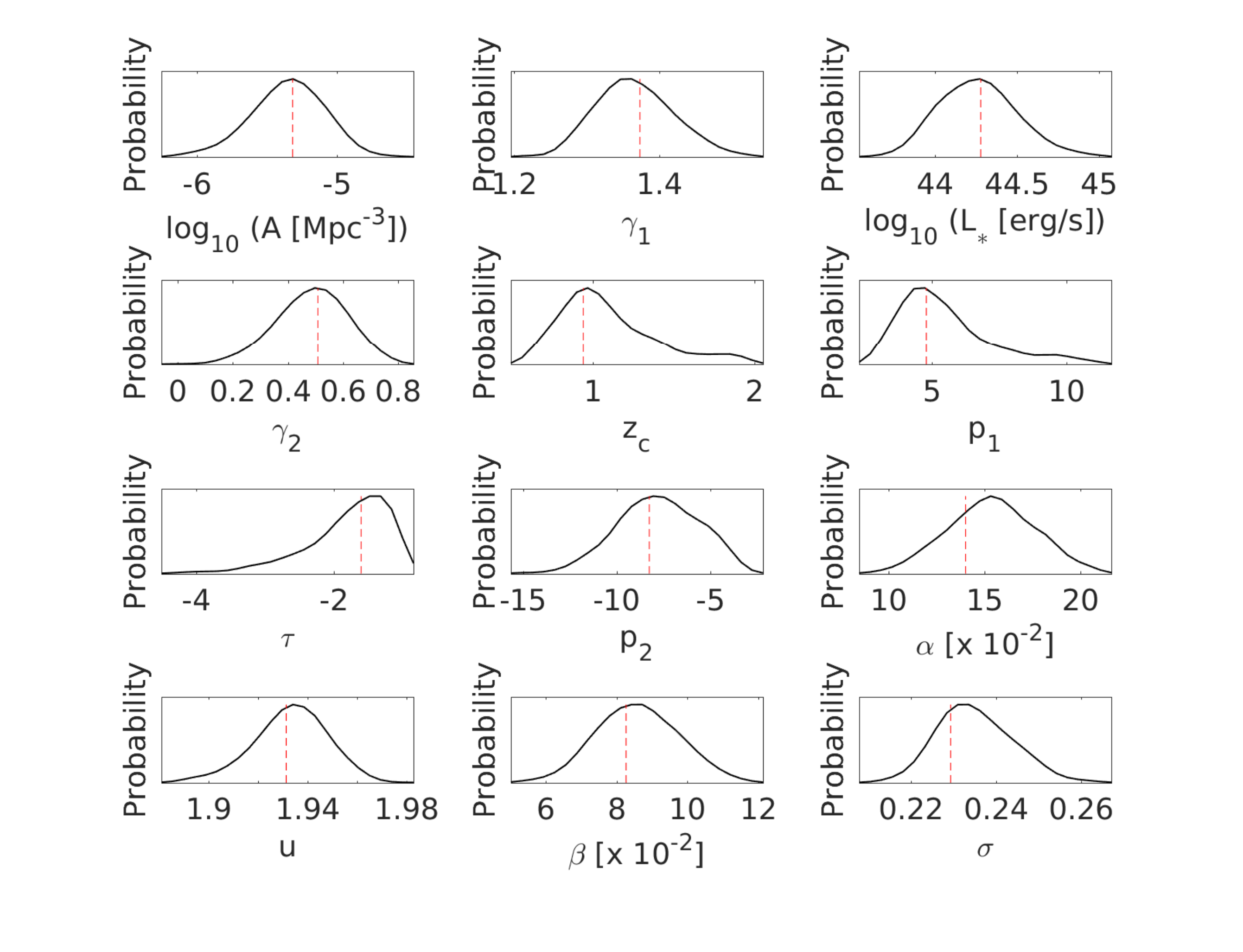}
\end{minipage}
}
\subfigure[LDDE]{
\begin{minipage}[t]{0.48\linewidth}
\centering
\includegraphics[width=1.0\linewidth]{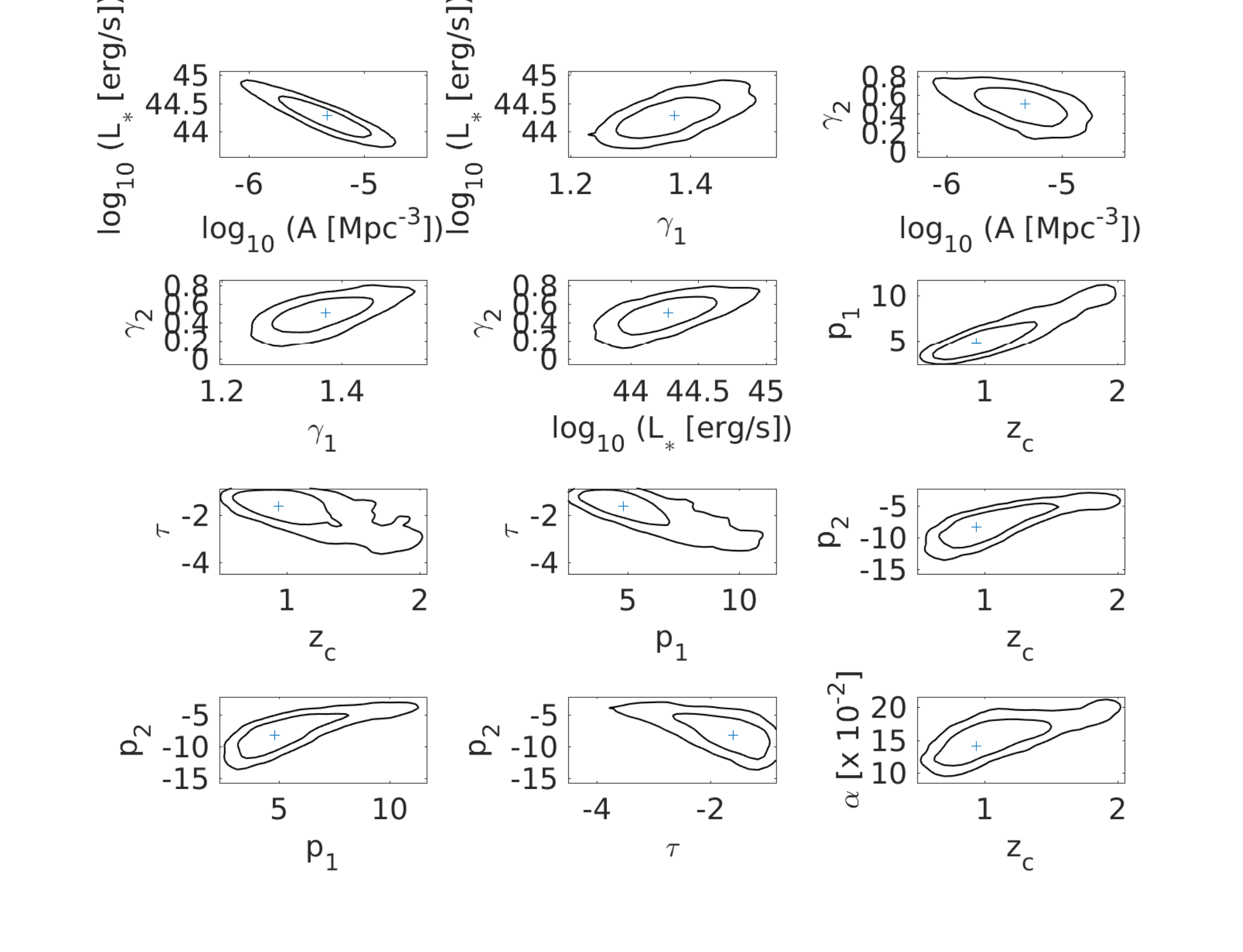}
\end{minipage}
}
\centering
\caption{ Left: The 1D probability distributions of the parameters and the best-fitted value (dashed line) for the (a) PDE, (c) PLE, and (e) LDDE models. Right: The 2D probability distributions of the parameters  for the (b) PDE , (d) PLE , and (f) LDDE models.}
\label{parameters}
\end{figure*}

\begin{figure*}
\centering
\subfigure{
\begin{minipage}[t]{0.45\linewidth}
\includegraphics[width=\linewidth]{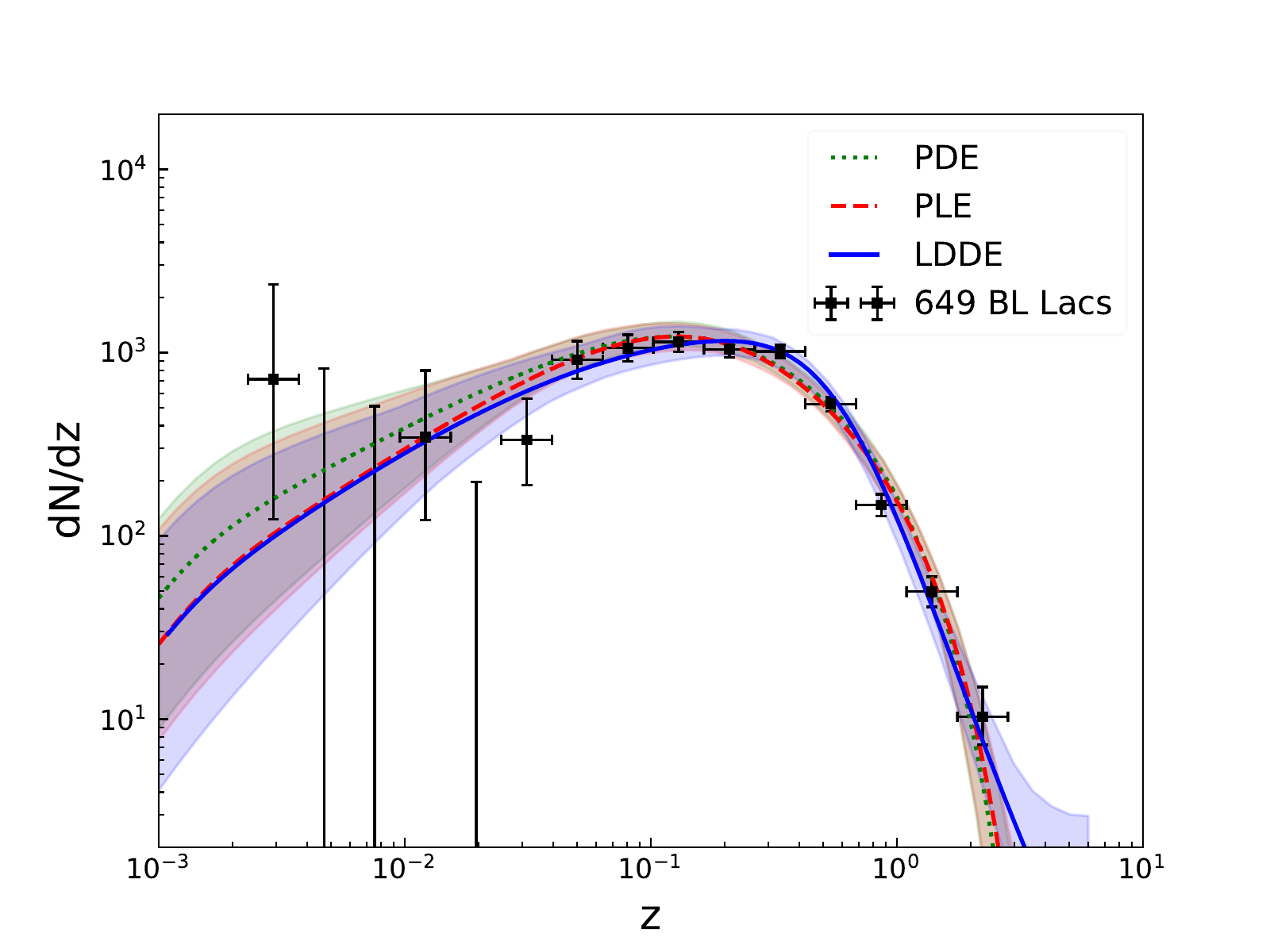}
\end{minipage}
}
\subfigure{
\begin{minipage}[t]{0.45\linewidth}
\includegraphics[width=\linewidth]{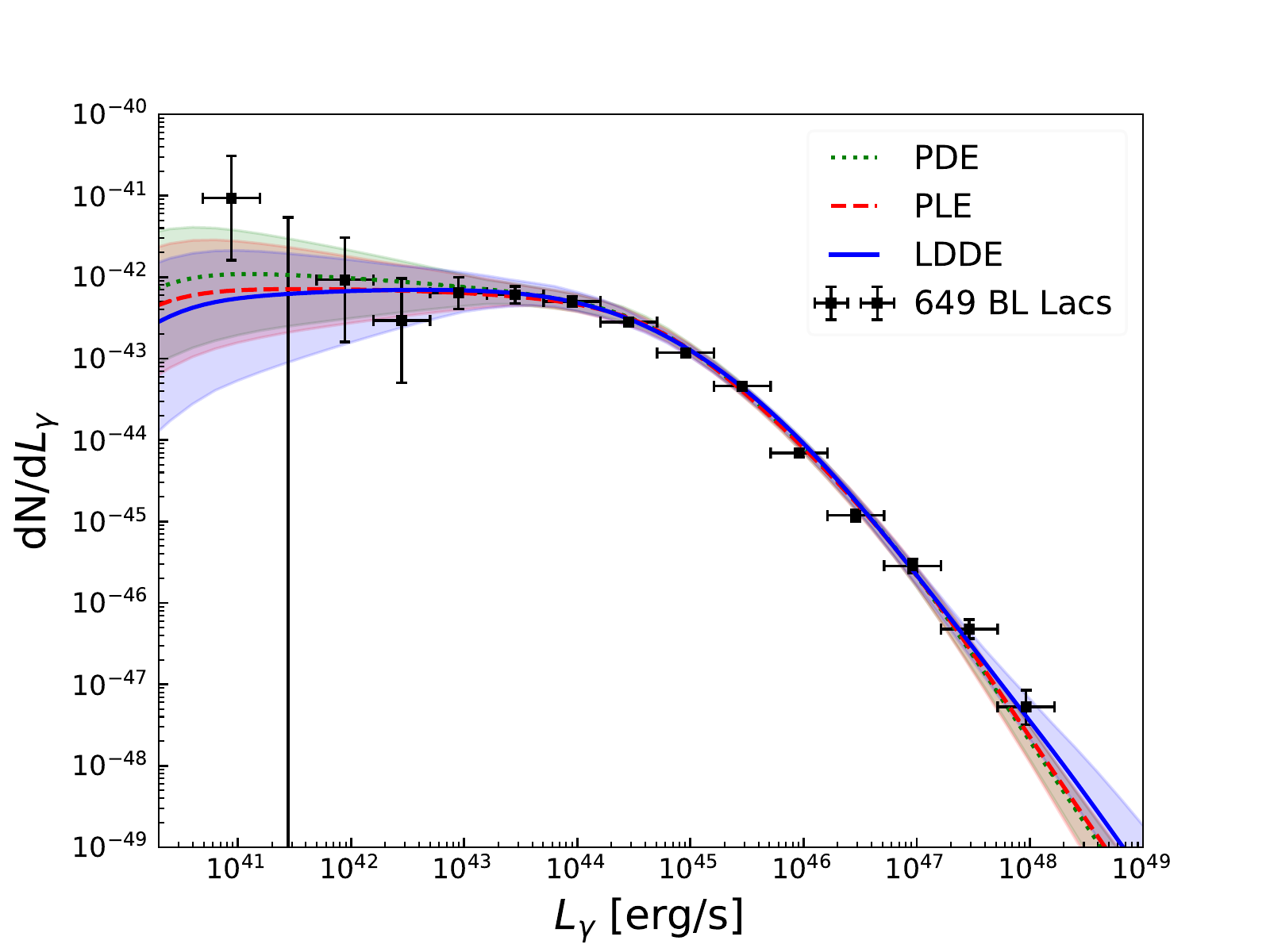}
\end{minipage}
}
\subfigure{
\begin{minipage}[t]{0.45\linewidth}
\includegraphics[width=\linewidth]{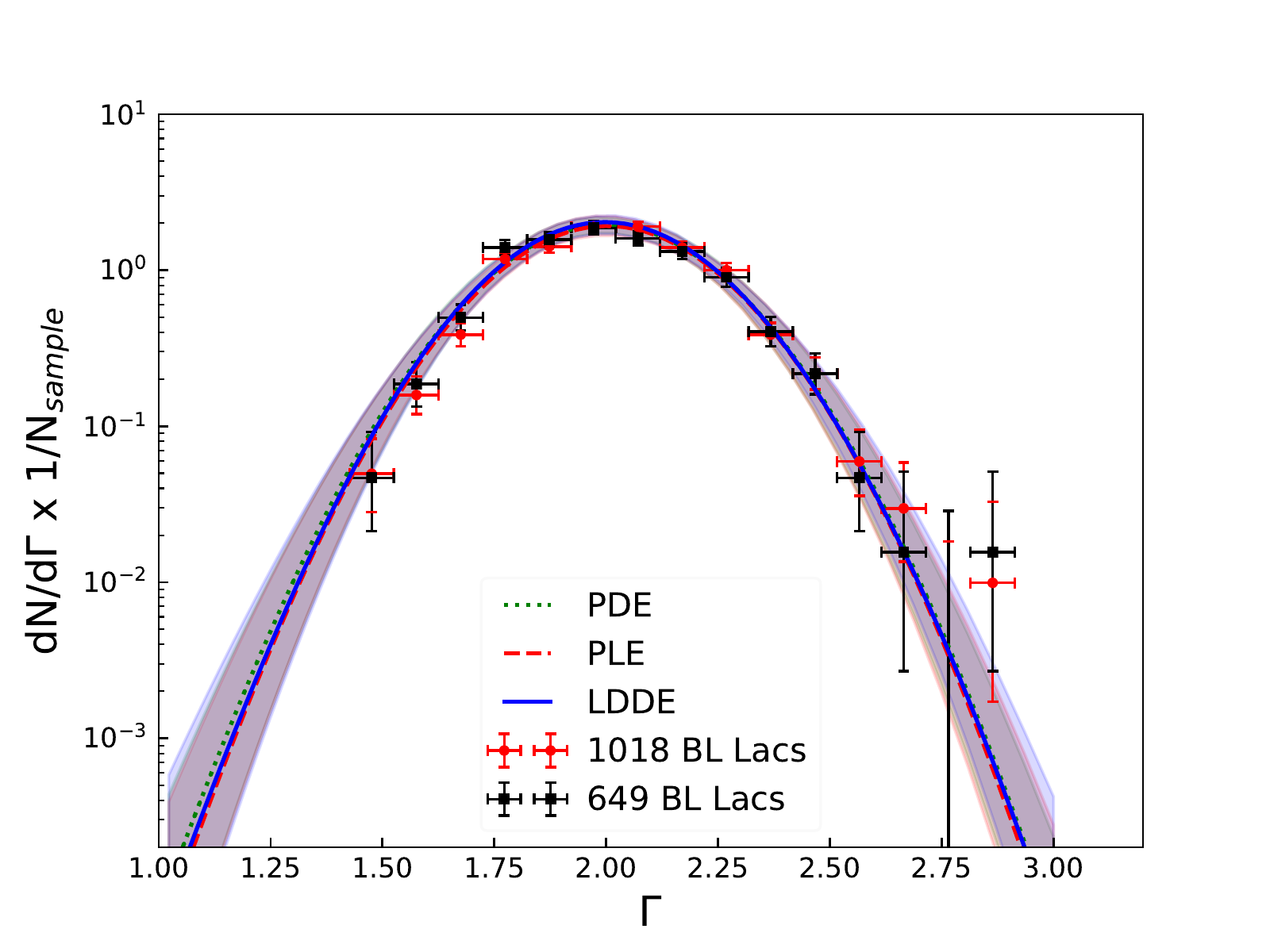}
\end{minipage}
}
\subfigure{
\begin{minipage}[t]{0.45\linewidth}
\includegraphics[width=\linewidth]{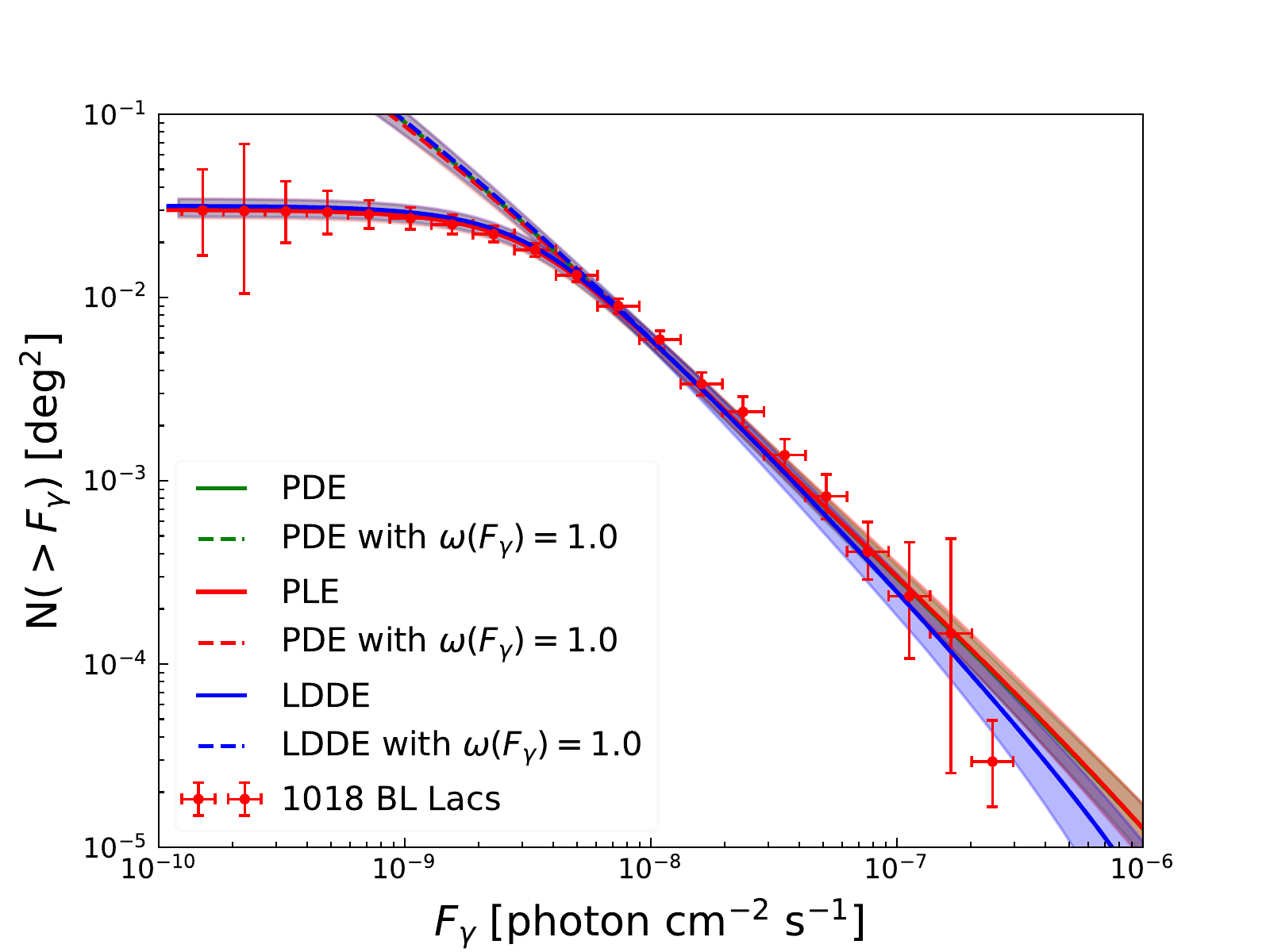}
\end{minipage}
}
\centering
\caption{ Redshift (upper left), luminosity (upper right), photon index
(lower left), and source count (lower right) distributions of BL Lacs. The
lines are the predictions of the best-fitting models. For photon index and source count distribution, a clean 4LAC sample (1018 BL Lacs) is shown in the red circle. The error bars represent
1$\sigma$ Poisson error}
\label{fitdistribution}
\end{figure*}

\begin{landscape}
\begin{table}
\renewcommand\arraystretch{1.0}
\caption{The best-fitting parameters of the GLFs of BL Lacs in PDE, PLE, and LDDE models. Errors are 1$\sigma$ statistical uncertainties.}
\label{tab:1}
\begin{tabular}{lcccccccccccc}
\hline
\hline
    Model &  ${\rm Log}_{10} A^{a} $ & $\gamma1$ & ${\rm Log}_{10}L_{\ast}$ & $\gamma2$  &   $  k_{\ast}$ or $p1^{\ast} $ & $ \tau $  & $ \xi$ or $p2 $  &$z_{c}$&$\alpha$& $ u $  & $ \beta $  & $ \sigma $ \\
    \hline
    PDE &
$-6.41$&
$  1.89$&
$ 44.69$&
$  0.72$&
$ 13.53$&
$  1.97$&
$ -0.13$&
&
&
$  1.93$&
$  0.080$&
$  0.23$\\
CL 68\%&$[-6.60,-6.22]$&
$[  1.75,  1.99]$&
$[ 44.55, 44.87]$&
$[  0.57,  0.75]$&
$[ 10.19, 14.17]$&
$[  1.39,  2.24]$&
$[ -0.17, -0.13]$&
&
&
$[  1.91,  1.94]$&
$[  0.063,  0.084]$&
$[  0.23,  0.24]$
\\
\hline
$S_{all}$ & $69387.3$ &&&&&&&&&&&\\
\hline
Sys errlow &$-6.25$&
$  2.05$&
$ 44.63$&
$  0.76$&
$ 16.46$&
$  1.87$&
$ -0.11$&
&
&
$  1.91$&
$  0.073$&
$  0.23$
\\
Sys errup &$-6.85$&
$  1.95$&
$ 45.09$&
$  0.71$&
$ 10.59$&
$  2.53$&
$ -0.15$&
&
&
$  1.95$&
$  0.073$&
$  0.23$
\\
\hline
\hline
PLE&$ -6.39$&
$  0.62$&
$ 44.72$&
$  1.92$&
$  4.23$&
$  0.70$&
$ -0.43$&
&
&
$  1.93$&
$  0.066$&
$  0.23$\\
CL 68\%&$[ -6.52, -6.11]$&
$[  0.53,  0.71]$&
$[ 44.49, 44.83]$&
$[  1.74,  1.97]$&
$[  3.51,  4.61]$&
$[  0.47,  0.72]$&
$[ -0.53, -0.39]$&
&
&
$[  1.91,  1.94]$&
$[  0.061,  0.083]$&
$[  0.22,  0.24]$\\
\hline
$S_{all}$ & $69389.5$ &&&&&&&&&&&\\
\hline
Sys errlow&$ -5.89$&
$  0.60$&
$ 44.39$&
$  1.96$&
$  5.54$&
$  0.53$&
$ -0.35$&
&
&
$  1.91$&
$  0.070$&
$  0.23$\\
Sys errup&$ -6.96$&
$  0.70$&
$ 45.23$&
$  1.96$&
$  3.08$&
$  0.83$&
$ -0.50$&
&
&
$  1.94$&
$  0.065$&
$  0.23$\\
\hline
\hline
  LDDE&
$ -5.32$&
$  1.37$&
$ 44.28$&
$  0.51$&
$ 4.81$&
$  -1.60$&
$  -8.27$&
$  0.94$&
$ 0.14$&
$ 1.93$&
$  0.083$&
$  0.23$\\
CL 68\%&$[ -5.61, -5.09]$&
$[  1.31,  1.42]$&
$[ 44.03, 44.49]$&
$[  0.36,  0.61]$&
$[ 3.93, 7.16]$&
$[  -2.36,  -1.23]$&
$[  -9.93, -5.42 ]$&
$[  0.81,  1.39 ]$&
$[  0.13,  0.18]$&
$[1.92, 1.95]$&
$[ 0.074,  0.098 ]$&
$[ 0.23,  0.24]$\\
\hline
$S_{all}$ & $69366.7$ &&&&&&&&&&&\\
\hline
Sys errlow&
$ -4.74$&
$  1.52$&
$ 44.10$&
$  0.55$&
$ 2.45$&
$  -1.03$&
$  -14.19$&
$  0.71$&
$ 0.13$&
$ 1.91$&
$  0.090$&
$  0.23$\\
Sys errup&
$ -5.75$&
$  1.21$&
$ 44.33$&
$  0.48$&
$ 6.60$&
$  -1.55$&
$  -6.35$&
$  1.38$&
$ 0.22$&
$ 1.96$&
$  0.084$&
$  0.23$\\

\hline
\hline
\end{tabular}

Notes.{Parameter values were the best-fit parameters to the Monte Carlo sample, CL \% represent the 68\% containment region around the median value.\\
Sys errlow and errup correspond to the results of using the detection efficiency (Figure .1) with the low- and up-end error, respectively. \\
$^{a}$  In unit of Mpc$^{-3}$.
 }
\end{table}
\end{landscape}

\subsection{Results}
Figure \ref{parameters} shows the one-dimensional (1D) probability distributions and the best-fitting values, and the two-dimensional (2D) probability distributions (1 $\sigma$  and 2 $\sigma$ levels) of the parameters for the three models.
It can be seen that all parameters of the three models are constrained well. The best-fitting values of the parameters are given in Table. \ref{tab:1}

Figure \ref{fitdistribution} shows the best fitting results to the observed data (i.e., redshift, luminosity, photon index, and source count distribution) of our BL Lac sample by using the three models.
The Akaike information criterion (AIC; \cite{1974ITAC...19..716A,2007MNRAS.377L..74L}),
${\rm AIC}_i = 2k +S_{\rm all}$ where $k$ is the number of model parameters and $S_{\rm all}$ is twice of the log-likelihood value reported in Table. \ref{tab:1}, is used to statistically determine
the model that is preferred by the data.
The relative likelihood value between two models can
be evaluated as $p = {\rm exp}(({\rm AIC}_{\rm min}-{\rm AIC}_i)/2)$ ,
where ${\rm AIC}_{\rm min}$ comes from
the model providing the minimal AIC value \citep[e.g.][]{Ajello_2013}.
The LDDE model has a relative likelihood with respect to the PDE model of $2.5 \times 10^{-4}$, and to the PLE model of $8.3 \times 10^{-5}$.
Namely, the LDDE model gives the better ($\sim 4\sigma$) fitting to the data than the PDE and PLE models, which agrees with the result of \cite{Ajello_2013}.

Comparing with the results obtained by \cite{Ajello_2013} and \cite{2014ApJ...786..129D}, a harder spectral index distribution ($\mu \sim 1.93$) is obtained,
implying that the most sources belongs to the high-synchrotron-peaked (HSP) BL Lacs.
The mean spectral index of HSP BL Lacs obtained by  \cite{Ajello_2013} and \cite{2014ApJ...786..129D} are 1.97 and 1.86, respectively.
The luminous HSP BL Lacs with hard GeV spectra are easily detected by \textit{Fermi}-LAT,
which are often estimated as high $z$ objects \cite[e.g.][]{2012A&A...538A..26R,2013ApJ...764..135S}.
Our results indicate that the break luminosity ($L_{\ast}$) of the three models are all $\sim 10^{44}$,
which is significantly lower than that ($\sim 10^{48}$) of \cite{Ajello_2013} and \cite{2014ApJ...786..129D} results, except for \cite{Ajello_2013}'s PLE model.

\begin{figure*}
\centering
\subfigure{
\begin{minipage}[t]{0.45\linewidth}
\includegraphics[width=\linewidth]{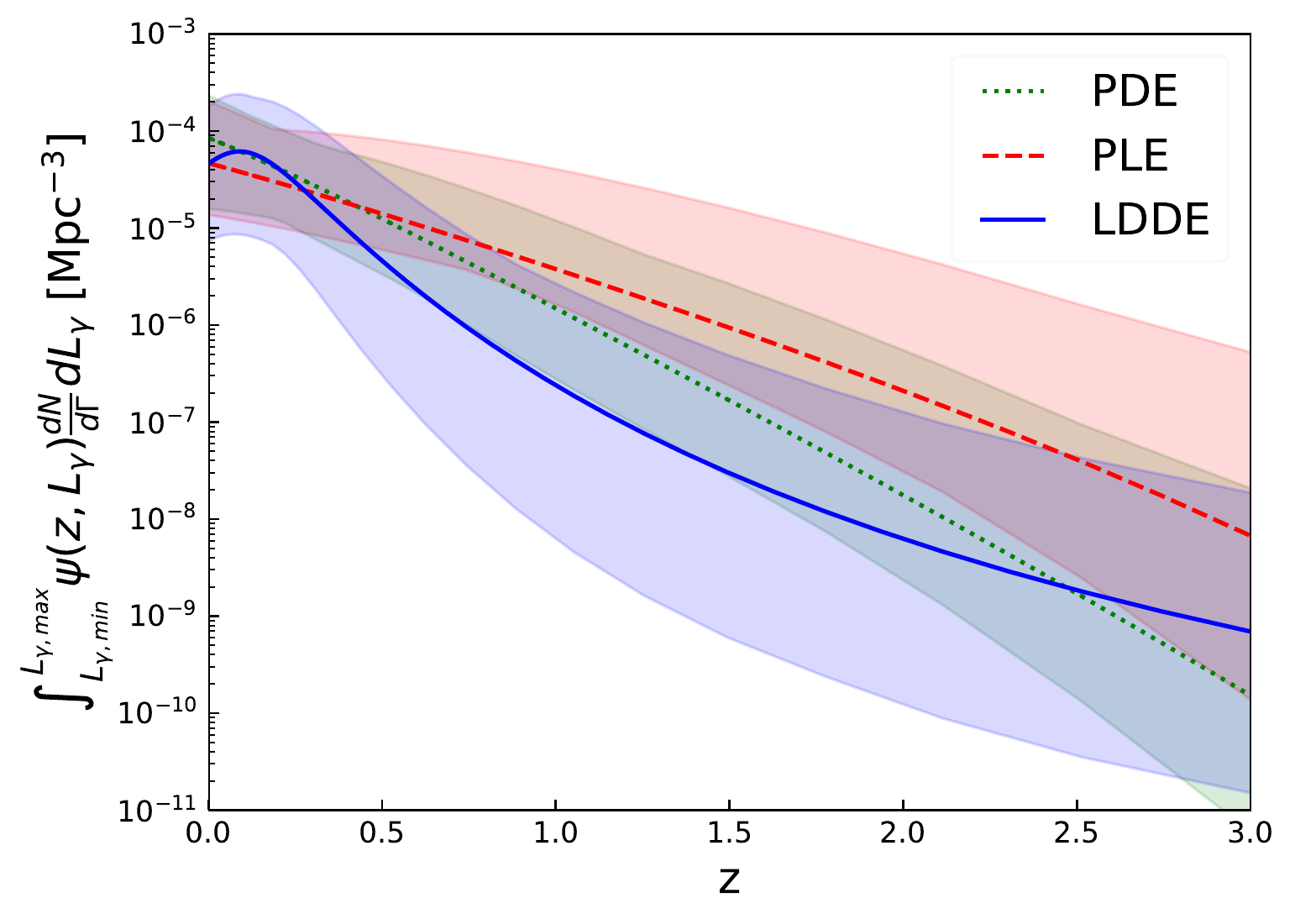}
\end{minipage}
}
\subfigure{
\begin{minipage}[t]{0.45\linewidth}
\includegraphics[width=\linewidth]{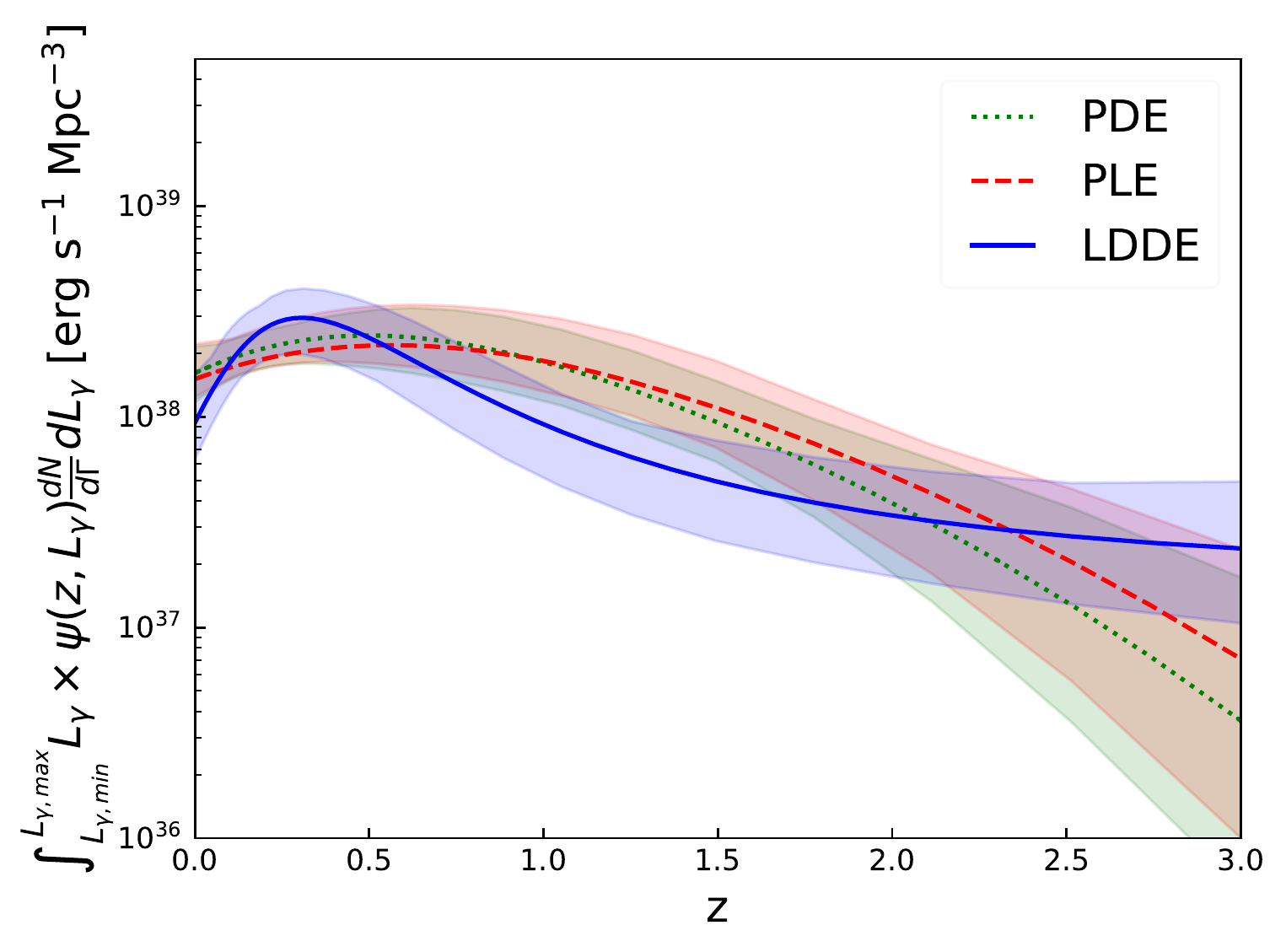}
\end{minipage}
}
\centering
\caption{ Left: Number density of BL Lac objects. Right: Luminosity density of BL Lacs. The clear band represents 68\% the confidence region.}
\label{Dedistribution}
\end{figure*}

Figure \ref{Dedistribution} shows the number density distribution (per unit comoving volume)
and the evolution of the luminosity density of BL Lacs, which do not show a complex evolution.
For the LDDE model, low $z$ BL Lacs ($z < z_c*(L_{\gamma}/10^{48})^{\alpha} \sim 0.25$ with $L_{\gamma}=10^{44}$ erg/s) show negative evolution with $p2=-8.27$ ,
which agrees with previous results,
and for $L_{\gamma} >10^{48}$ erg/s, the redshift peak is $\sim$ 1.0.
High $z$ BL Lacs present a positive evolution with $p_1=4.81-1.60({\rm{log}_{10}}L_{\gamma}-46)$
The evolution slows down with increasing luminosity, and becomes non-evoluton
station when $L_{\gamma} \sim 10^{49}$, which is different from the result of
\cite{Ajello_2013}, but consistent with the result of \cite{zeng2014}.


\section{The contribution to EGB and IGRB}

The integration of the product of $F_{\gamma}$ and the space density distribution of BL Lacs over redshift, luminosity, spectral index is the energy-differential EGB,
\begin{equation}
\frac{dN}{dEd\Omega}=\int_{\Gamma_{min}}^{\Gamma_{max}}\int_{z_{min}}^{z_{max}}\int_{L_{\gamma min}}^{L_{\gamma max}}F_{\gamma}(L_{\gamma},z,\Gamma)\frac{d^{3}N}{dzdL\gamma d\Gamma}dzd\gamma dL_{\gamma}
\label{dn}
\end{equation}

Integrating Eq \ref{dn} over energy, we can get the contribution of BL Lacs to EGB,

\begin{equation}
I_{\rm EGB}=\frac{dN}{d\Omega}=\int_{E_{1}}^{E_{2}}\frac{dN}{dEd\Omega} dE
\end{equation}
If we only integrate the unsolved sources, in other words, multiplying a factor of $(1-\omega_{\gamma})$ in the integral operation,
we can get the contribution of unsolved BL Lacs to IGRB.

Based on an observational fact that blazars have curved GeV spectra,
we use a double power-law model \cite[e.g.][]{2015ApJ...800L..27A} for $F_{\gamma}$,
\begin{eqnarray}
\label{eq:Fg}
F_{\gamma}= N_0 \cdot \textrm{e}^{-\tau(E,z)}
 \left\{
\begin{array}{lcl}
E^{-\gamma_a}&
& { \textrm{if}~~~ E < E_{\rm br} } \\
 E_{\rm br}^{(\gamma_b-\gamma_a)} E^{-\gamma_b}  &
& { \textrm{if}~~~E \geq E_{\rm br} }\;,
\end{array}
\right.
\end{eqnarray}
where $\gamma_a=\Gamma$ , $\gamma_b$ and $E_{\rm br}$ are free parameters.
Here, $\tau(E,z)$ is the optical depth of gamma-rays traveling in extragalactic background light (EBL),
and we adopt the model of \cite{2010ApJ...712..238F}.
Note that the normalizing parameter $N_0$ is determined by Eq. \ref{Eq.2} with a simple power-law spectrum.

\begin{figure}
\centering
\subfigure[PDE]{
\begin{minipage}[t]{\linewidth}
\centering
\includegraphics[width=\linewidth]{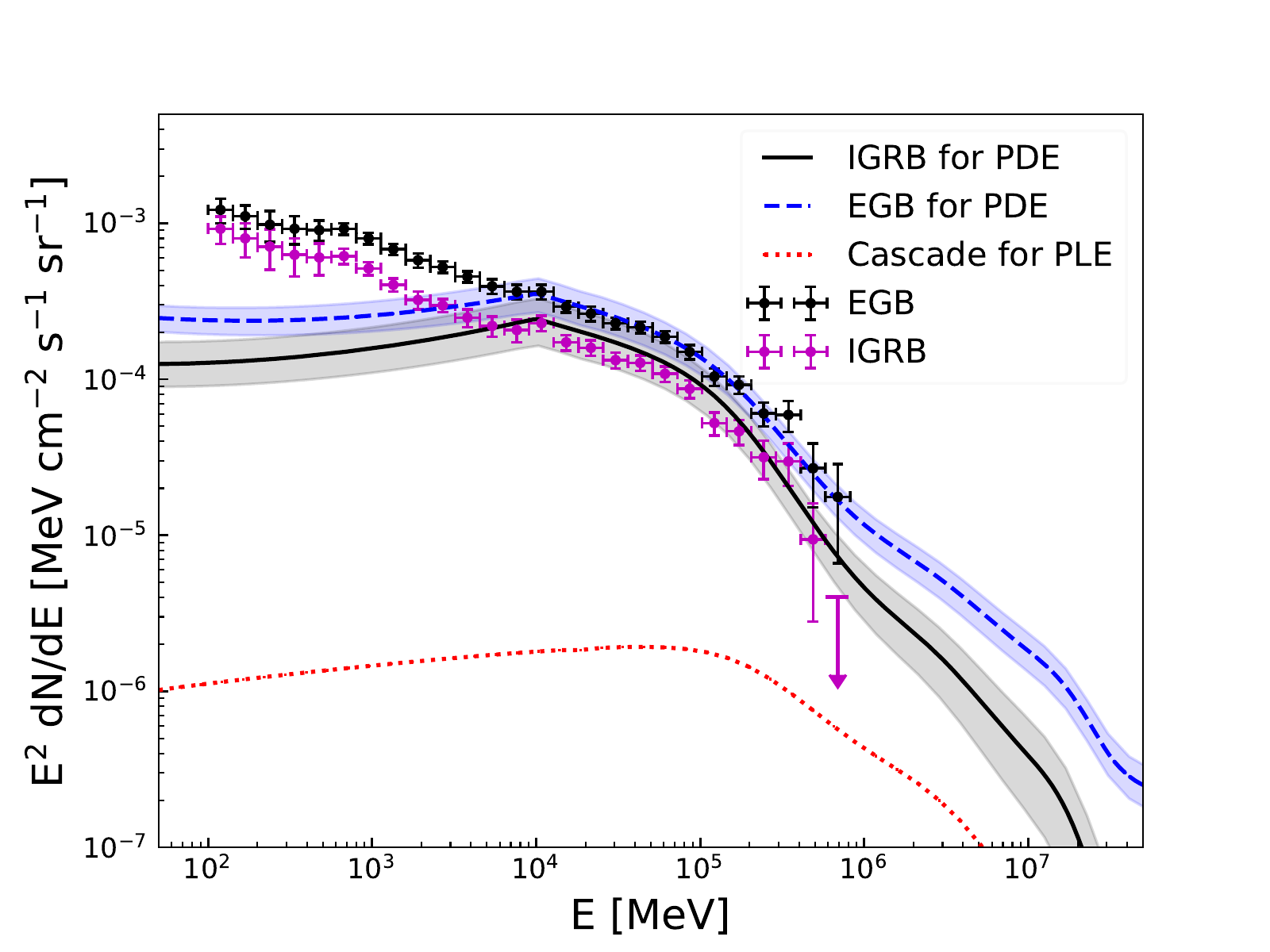}
\end{minipage}
}

\subfigure[PLE]{
\begin{minipage}[t]{\linewidth}
\centering
\includegraphics[width=\linewidth]{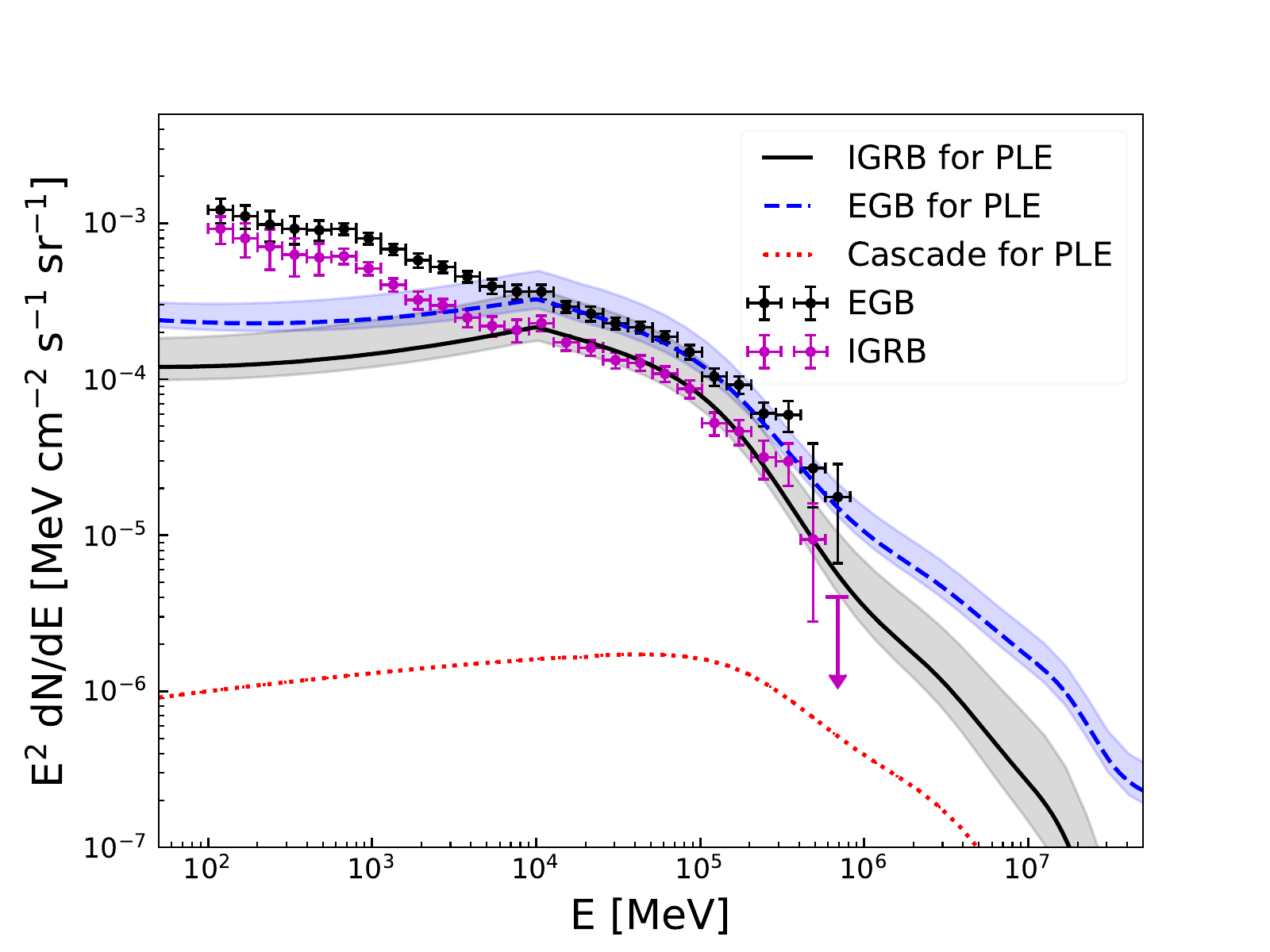}
\end{minipage}
}

\subfigure[LDDE]{
\begin{minipage}[t]{\linewidth}
\centering
\includegraphics[width=\linewidth]{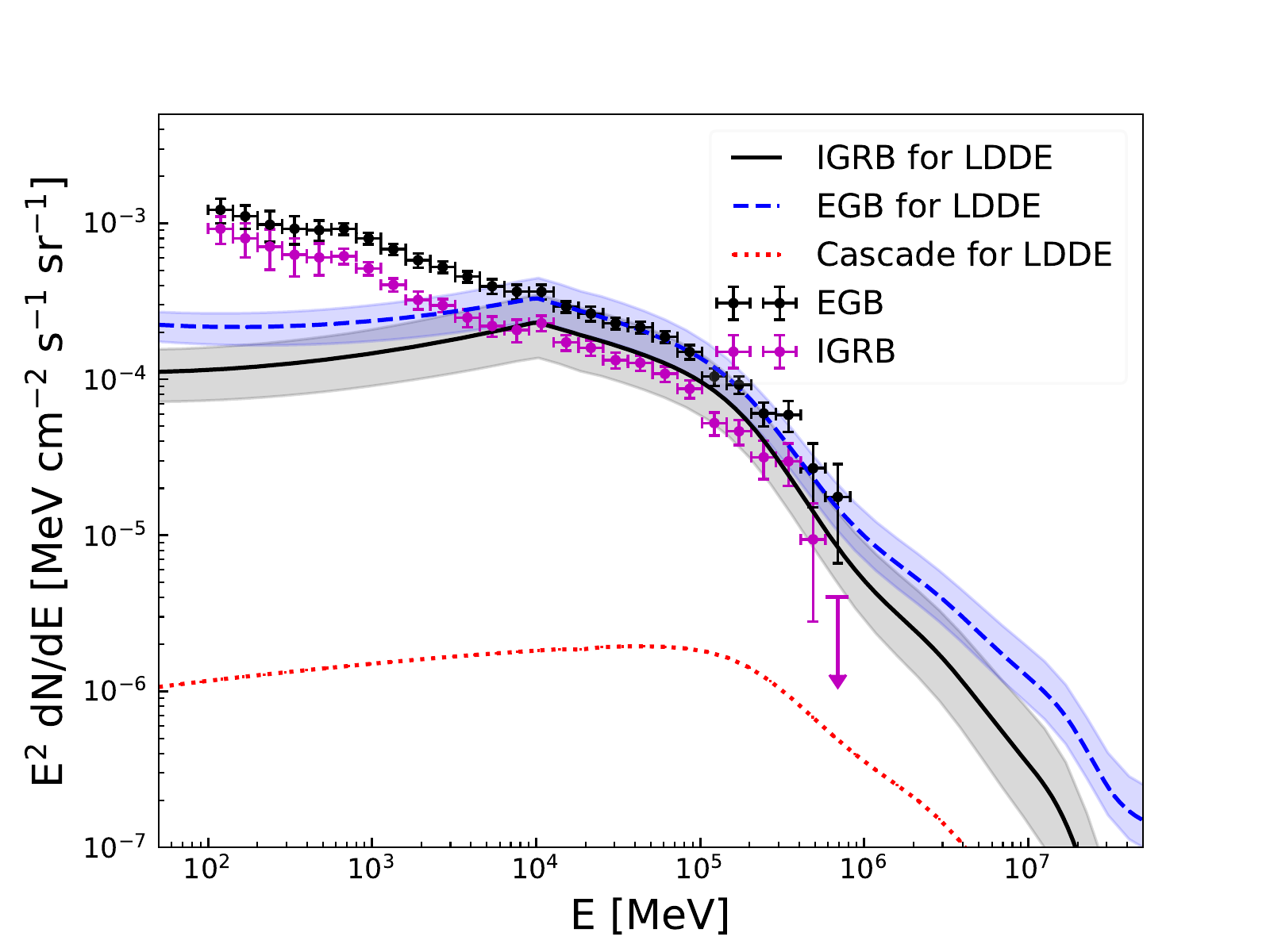}
\end{minipage}
}
\centering
\caption{Contribution of BL Lacs to EGB (dashed line), IGRB(solid line) calculated by the best-fitting parameters of PDE, PLE and LDDE models. The red line represents the estimated intensity of the cascade emission from BL Lacs.}
\label{fig:EGRB}
\end{figure}

Figure~\ref{fig:EGRB} shows the contributions from BL Lacs to EGB and IGRB spectra in different models.
In the calculations, we use $\gamma_{\rm b}=2.3$ and $E_{\rm br}=10 $ GeV.
 The predicted EGB intensities for $E> 100$ MeV and $E>50$ GeV contributed by the BL Lacs are respectively $\sim$ 20\% and $\sim $ 100\% of
 the intensity measured by \textit{Fermi}-LAT \footnote{ The measured EGB intensity is $1.13(\pm 0.17) \times 10^{-5}$ ph  cm$^{-2}$ s$^{-1}$ sr$^{-1}$ above 100 MeV and $2.37(\pm 0.36) \times 10^{-9} $ ph  cm$^{-2}$ s$^{-1}$ sr$^{-1}$ above 50 GeV, and the measured IGRB intensity is $7.2(\pm 0.6) \times 10^{-6}$ ph  cm$^{-2}$ s$^{-1}$ sr$^{-1}$ above 100 MeV \citep{Ackermann_2015}}.
 The contribution to IGRB is about 20\% of the measured intensity.
 The specific values are listed in Table. \ref{tab:2}.
 Those results are slightly higher that of \cite{Ajello_2013}, \cite{2014ApJ...786..129D} and \cite{PhysRevLett.116.151105}.
 Note that the EGB intensity with $E>50$ GeV depends heavily on the value of parameters $\gamma_{\rm b}$ and $E_{\rm br}$.
 In addition, we compute the contribution of the cascade emission of BL Lacs to the EGB \cite[e.g.][]{2012MNRAS.422.1779Y,zeng2014},
 and find that the contribution is negligible due to intrinsic curved gamma-ray spectra.

Our estimated EGB spectra extends to $\sim$10 TeV.
The future observation of the TeV extragalactic emission background by CTA would put stronger constraint on the evolution of BL Lacs.

\begin{table}
\renewcommand
\arraystretch{1.0}
\caption{The cotribution of BL Lacs to EGBs and IGRBs calculated by three different model}
\label{tab:2}
\begin{tabular}{lccc}
\hline
Model &  EGB($>100$MeV) & EGB$^{a}$($>50$ GeV)  & IGRB($>100$ MeV) \\
&$10^{-6} $ ph  cm$^{-2}$& $10^{-9} $ ph  cm$^{-2}$ & $10^{-6} $ ph  cm$^{-2}$\\
\hline
PDE &
$\sim 2.52$&
$\sim 2.64$&
$\sim 1.44$\\
&($\sim 22.3\%$)&($\sim 100\%$)&($\sim 20.0\%$)
\\
\hline
PLE&$\sim 2.42$
&$\sim 2.39$
&$\sim 1.35$
\\
&$ (\sim 21.4\%)$&
$  (\sim 100\%)$&
$ (\sim 18.8\%)$\\
\hline
  LDDE&$\sim 2.31$
&$\sim 2.58$
&$\sim 1.32$
\\
&$ (\sim 20.4\%)$&
$  (\sim 100\%)$&
$ (\sim 18.3\%)$\\

\hline
\end{tabular}
$^a$ the EGB intensity with $E>50$ GeV significant depends on the value of parameters $\gamma_b$ and $E_{\rm br}$
\end{table}

\section{Conclusions and Discussions}

In this paper, using an enlarged \textit{Fermi}-LAT BL Las sample, we first combine the space density distribution and source counts distribution to
construct the $\gamma$-ray luminosity function of BL Lacs with three commonly used forms (i.e., PLE, PDE, LDDE).
The combination of the two aspects can effectively reduce the sample incompleteness and obtain the {\it true} GLF.
The MCMC technique is applied for different models to
obtain the best-fitting evolutionary parameters.
The final fitting results show that the LDDE model can give best
description for BL Lac GLF than the other two models.
The best-fitting LDDE model shows that
the BL Lacs with a hard GeV spectrum evolve as strongly as FSRQs, and the evolution decreases as the increasing luminosity ($\tau=-1.60$).

Secondly, with the improved GLFs, we estimate the contribution of BL Lacs to EGB and IGRB.
Based on the result of \cite{PhysRevLett.116.151105} that the contribution of 2FHL sources to the EGB is close to 100 \%,
we constrain the mean photon spectra of BL Lacs with a double power law model.
The break energy is about 10 GeV and the high-end slope is about 2.3, which is harder than the slope obtained by \citet{2015ApJ...800L..27A} (2.6) and by \citet{Di_Mauro_2018} (2.8).
Our results show that BL Lacs contribute $ \sim 20\%$ of EGB at $E>100$ MeV and contribute $\sim 100\%$ of EGB at $E>50$GeV.
And the unresolved BL Lacs contribute about $20\%$ of IGRB at $E>100$ MeV.
Those result is slight higher than that of previous researches.
Predictions of TeV EGB and IGRB are also showed in Figure 5 which could be tested by future detectors.

\citet{2012ApJ...751..108A} suggested that the contribution from \textit{Fermi}-LAT FSRQs is $\sim9$\% of the IGRB intensity
in the 0.1-100 GeV band.
\citet{2019arXiv190306544S} found that the core dominated radio galaxies also contribute a small account ( $4\%$-$18\%$) of the IGRB.
\citet{2017PhRvD..96h3001L} claimed that star-forming galaxies significantly
contribute to the IGRB, producing $\sim61\%$ of the total IGRB intensity above 1 GeV.
According to these results, the entire observed IGRB can reasonably be explained by the contributions from blazars and star-forming galaxies.
In other words, the contribution from real diffuse processes (e.g., the decay or annihilation of dark matter and interactions of cosmic rays with background photons) is negligible.


\section*{Acknowledgements}
We are grateful to the anonymous referee for insightful comments. We acknowledge financial supports from  National Key R\&D Program
of China (2018YFA0404203), the National
Natural Science Foundation of China (NSFC-11703094, NSFC-U1738124 and NSFC-11803081) and the joint foundation of Department of Science and Technology of Yunnan Province and Yunnan University [2018FY001(-003)].
The work of D. H. Yan is also supported by the CAS
``Light of West China'' Program and Youth
Innovation Promotion Association.






\appendix




\bsp	
\label{lastpage}
\end{document}